\documentstyle[prl,aps,preprint,axodraw,12pt]{revtex}
\newcommand{\beqq}{\begin{equation}}
\newcommand{\eeqq}{\end{equation}}
\newcommand{\beq}{\begin{eqnarray}}
\newcommand{\beas}{\begin{eqnarray*}}
\newcommand{\eeq}{\end{eqnarray}}
\newcommand{\eeas}{\end{eqnarray*}}
\newcommand{\ba}{\begin{array}}
\newcommand{\ea}{\end{array}}
\begin{document}
\draft
\title{Electric charge 
quantization and the muon anomalous magnetic moment}
\author{ C. A de S. Pires and P. S. Rodrigues da Silva}
\address{\tightenlines{\it Instituto de  F\'{\i}sica 
Te\'{o}rica, Universidade 
Estadual Paulista.\\ Rua Pamplona 145, 01405-900 S\~{a}o Paulo, 
S\~{a}o Paulo, Brazil.}}
\date{August, 2001} 
\maketitle
\begin{abstract}
We investigate some proposals to solve the electric charge 
quantization puzzle, which simultaneously explain the recent 
measured deviation on the muon anomalous magnetic moment. For 
this we assess extensions of the Electro-Weak Standard Model 
spanning modifications on the scalar sector only. It is 
interesting to verify that one can have modest extensions which 
easily account for the solution for both problems. 
\end{abstract}

\section{introduction}
It is known that the Minimal Standard Model (MSM), though very 
well tested at the experimental level, is not the most complete 
theory of particle physics since some important questions cannot 
be explained without quoting physics beyond its minimal 
structure. Among these questions there is an intriguing one which 
concerns the observation of, to an extremely high accuracy, exact 
electric charge quantization (ECQ). The fact that differences 
among electric charges of known particles are given in terms of 
integer numbers, implies that any reasonable theory for the 
elementary particles has to accommodate such a quantization or, 
at least, must give a strong reason for the violation to be so 
small. 

In the beginning of the nineties, the question concerning  ECQ 
was studied inside the MSM through classical and quantum 
constraints~\cite{footjoshi,babumoh1,zralek,babumoh2,geng,marek,babumoh3,pires}.
The classical constraints come from imposing invariance of the 
Lagrangian under the standard gauge group transformation, 
$SU(3)_C \otimes SU(2)_L \otimes U(1)_Y$, while the quantum 
constraints are consequence of anomaly cancellation via the 
computation of the triangle diagrams\cite{jackiw}. If these two 
constraints fix the pattern of quantization of hyper-charge, they 
automatically establish the pattern of quantization for the 
fermion electric charge through the formula, 
\beqq Q=T^3 +\frac{Y}{2}\,, \label{e0} \eeqq
where $T^3$ is the diagonal generator of $SU(2)$ and $Y$ is the 
hyper-charge of the particle with charge $Q$. All sectors of the 
MSM Lagrangian are trivially gauge invariant by the standard 
gauge symmetry, except the Yukawa sector. It is enough to focus 
only on this part of the Lagrangian wherein the imposition of 
gauge invariance implies the useful classical relations among the 
hypercharges of fermions and scalars, still arbitrary at this 
level. This formalism can be extended to any model based on 
semi-simple group with structure $SU(3)_C \otimes SU(n)_L \otimes 
U(1)_X$. In this case, the constraints must fall over the quantum 
number $X$, and the electric charge is given by a combination of 
$X$ and the diagonal generators, $T_d$, of $SU(n)$. Then, instead 
of Eq.~(\ref{e0}) we get,
\beqq Q=\sum T_d +\frac{X}{2}\,. \label{e01} \eeqq

It is interesting to remind that MSM with only one family has 
enough content to provide ECQ\cite{footjoshi,babumoh1}. However, 
when three families are considered this property is lost and the 
theory undergoes an effect known as electric charge 
dequantization. The reason is that with three families, 
combinations of additional $U(1)$ gaugeable symmetries start to 
plague the model in the sense that, besides hyper-charge, one has 
now a continuous amount of extra quantum numbers. These cannot be 
fixed by the present constraints on MSM, forbidding ECQ. Then, we 
can state that if the formalism is applied to extensions of MSM 
and predict the ECQ, this means the model has no other global 
symmetry that can be promoted to local symmetry besides 
$U(1)_Y$\cite{footjoshi,babumoh1}. Conversely, if there exists 
some global symmetry, potentially ``gaugeable'', usually called 
hidden symmetry, we cannot obtain enough classical or quantum 
constraints to have ECQ. Unfortunately, this is the case if one 
sticks to the MSM alone\cite{footjoshi,babumoh1,zralek,pires},
 where there exist three 
hidden symmetries (when mixed gauge-gravitational anomalies are 
also considered): $U(1)_{L_e-L_\mu , L_e-L_\tau ,L_\mu-L_\tau}$, 
which are the gaugeable lepton flavor symmetries. This implies 
that electric charge in the MSM cannot be quantized and has the 
following expression: $Q = Q_{st} +\epsilon (L_e-L_\mu\,\, 
\mbox{or}\,\, L_e-L_\tau \,\, \mbox{or}\,\, L_\mu-L_\tau 
)$\cite{footjoshi,babumoh1}, 
where $Q_{st}$ is the standard assigned charge to the respective 
particle and  $\epsilon$ is an arbitrary continuous parameter. 

This approach paved the way one needs to follow in order to 
suitably construct models beyond MSM, to naturally obtain ECQ: 
extensions of MSM that aim to predict ECQ must explicitly break 
any hidden symmetry at the Lagrangian level. In this direction 
various  extensions of the MSM were analyzed in 
literature~\cite{footjoshi,zralek,babumoh2,marek,pires}. The main conclusion 
achieved on those works 
was that if neutrinos are massive and Majorana-like, then they 
automatically yield ECQ~\cite{babumoh2}. This is easy to 
understand since  Majorana mass terms  do require violation of 
the lepton number and, as a consequence, the hidden symmetries  
$U(1)_{L_e-L_\mu , L_e-L_\tau ,L_\mu-L_\tau} $ must be explicitly 
broken. This idea received a great deal of attention at that time 
because there was experimental evidence of neutrino oscillations, 
whose solution demands neutrinos to be massive. Recently such 
oscillations were confirmed, however it is  not possible to infer 
the true nature of neutrinos from those experiments, if Majorana 
or Dirac, leaving unsolved the ECQ puzzle~\cite{foot}.

Nevertheless, a new measurement of the anomalous magnetic moment 
of the muon, $(g-2)_\mu$, has shown a deviation from the 
theoretical result~\cite{brown}, pointing to possible new physics 
beyond the MSM. Since the deviation is in the leptonic sector, it 
turns out that its solution can, conveniently, be cast in such a 
way to simultaneously solve the ECQ puzzle. This is what we 
propose in this work. In order to do that, we first review the 
casting of ECQ in the MSM at section~\ref{sec1}. Then we present, 
at section~\ref{sec2}, simple MSM extensions suitable to 
correctly get the ECQ and show, in section~\ref{sec3}, that some 
of these proposals can explain the $(g-2)_\mu$ deviation. 
Finally, we present our conclusions in section~\ref{sec4}.

\section{ECQ within Standard Model}
\label{sec1}

In order to better understand the method employed to study the 
ECQ, we start by reviewing the procedure in the context of the 
MSM. The MSM is defined by the gauge structure $SU(3)_C \otimes 
SU(2)_L \otimes U(1)_Y$. According to this structure, let us 
attribute the following representation to the fermions,
\begin{eqnarray}
&&L_{i_L}= \left (
\begin{array}{c}
\nu \\
e
\end{array}
\right )_L \,\, \sim (\mbox{{\bf 
1,2}},Y_{i_L}),\,\,\,\,\,\,l_{i_R} \sim 
(\mbox{{\bf 1, 
1}},Y_{i_l}),\nonumber \\
&&Q_{i_L}=
\left (
\begin{array}{c}
u \\
d
\end{array}
\right )_L \,\, \sim (\mbox{{\bf 3}}, \mbox{{\bf 
2}},Y_{i_Q}),\,\,\,\, u_{i_R} \sim (\mbox{{\bf 
3,1}},Y_{i_u}),\,\,\,\, d_{i_R} \sim (\mbox{{\bf 
3,1}},Y_{i_d})\,,  \label{rep1} 
\end{eqnarray} 
where the index $i=1,\,2,\,3$ labels the three different families 
of leptons, $L_{i_L}$ and $l_{i_R}$,  and quarks, $Q_{i_L}$, 
$u_{i_R}$ and $d_{i_R}$. The sub-indexes $L$ and $R$ stand for 
left-handed and right-handed projections. The Yukawa sector of 
the MSM is given by,
\begin{equation}
{\cal L}^Y=g^l_{ii} \bar{L}_{i_L} \phi l_{i_R} + g^u_{ij} 
\bar{Q}_{i_L} \tilde \phi u_{j_R} +g^d_{ij} \bar{Q}_{i_L} \phi 
d_{j_R} + h.c.\,, \label{yukif}
\end{equation}
with $g^l_{ii},\, g^u_{ij}$ and $g^d_{ij}$\cite{zralek}, the usual Yukawa 
couplings. Under the ${\rm U}(1)_Y$ gauge invariance, 
Eq.~(\ref{yukif}) gives us the following constraints,
\begin{equation}
Y_{i_l}=Y_{i_L} -1,\,\,\,\,\, Y_{j_u}=Y_{i_Q} +1,\,\,\,\,\, 
Y_{j_d} =Y_{i_Q} -1\,. \label{clasconst1}
\end{equation}
The last two constraints amount to,
\beq
 && Y_{1_u}=Y_{2_u}=Y_{3_u}=Y_{u},\nonumber \\
 && Y_{1_d}=Y_{2_d}=Y_{3_d}=Y_{d},\nonumber \\
 && Y_{1_Q}=Y_{2_Q}=Y_{3_Q}=Y_{Q}, 
 \label{rel1}
 \eeq
leaving us with the true constraints, 
\begin{equation}
Y_{i_l}=Y_{i_L} -1,\,\,\,\,\, Y_{u}=Y_{Q} +1,\,\,\,\,\, Y_{d} 
=Y_{Q} -1.
\label{clasconst2}
\end{equation}
It is clear from the equations above that we have 5 free 
parameters. In order to fix these parameters we need additional 
equations which can be taken from the anomaly cancellation 
constraints. However, once Eq.~(\ref{clasconst2}) is taken into 
account, the MSM presents only two non-trivial anomalies whose 
vanishing condition furnishes two more constraints over the 
hyper-charges, 
\begin{eqnarray}
&&[{\rm SU}(2)_L]^2 {\rm U}(1)_Y\Longrightarrow 
9Y_Q+\sum_{i}^{3}Y_{L_i}=0 
,\nonumber \\
&&[{\rm U}(1)_Y]^3_Y \Longrightarrow 
18Y^3_Q-9Y^3_u-9Y^3_d+\sum_{i}^{3}(2Y^3_{i_L}-Y^3_{i_L})=0\,, 
\label{anomcanc}
\end{eqnarray}
and these, together with Eq.~(\ref{clasconst2}), are not enough 
to fix all the  hypercharges. In short, this is so because the 
leptonic sector is not as constrained as the quark sector, and 
presents extra global symmetries which can be promoted to gauge 
symmetries, namely, $U(1)_{L_e-L_\mu , L_e-L_\tau ,L_\mu-L_\tau} 
$. As was remarked in the introduction, the presence of these 
gaugeable symmetries forbids ECQ for the MSM. 

In view of this, we can say that MSM lacks additional constraints 
in the leptonic sector once ECQ is realized in nature. If one 
focuses on this issue only, it is clear that appropriate 
extensions of MSM have to be related mainly to the leptonic 
sector. Moreover, they should include terms that explicitly 
forbid the above hidden symmetries, automatically reducing the 
number of free leptonic hyper-charges to only one, i.e., the new 
terms ought to provide the following relations among such 
hyper-charges,
\beq \left.
\begin{array}{cccc}
  Y_{1_L}= & Y_{2_L}= & Y_{3_L}= & Y_{L} \\
  Y_{1_l}= & Y_{2_l}= & Y_{3_l}= & Y_{l} 
\end{array}
\,\,\right\}\,\,\,\Longrightarrow\,\,\,Y_l=Y_L -1\,. 
\label{sistem2} \eeq 
Substituting this result in Eq.~\ref{clasconst2}, yields,
\beq Y_l=Y_L -1,\,\,\,\,\, Y_{u}=Y_{Q} +1,\,\,\,\,\, Y_{d} =Y_{Q} 
-1\,, \label{clasconst4} \eeq
which, along with the anomaly cancellation constraints in 
Eq.~(\ref{anomcanc}), are enough to fix the hyper-charges, 
\[
Y_L = -1,\,\, Y_l=-2,\,\, Y_Q=1/3,\,\, Y_u=4/3,\,\, Y_d=-2/3\,. 
\]
This leads automatically to the ECQ with the correct electric 
charge pattern, $Q_l =-1,\,\,Q_u=1/3,\,\,Q_d=-2/3$.

Observe that we can be driven to the relations in 
Eq.~(\ref{sistem2}) from operators involving  bilinear fermionic 
products like $\bar \Psi^c_i \Psi_j$. These operators violate 
twice the total fermion number and the simplest kind of particle 
that can couple to this sort of bilinear, embedded  in a small 
extension of the MSM, are scalars carrying $F=-2$. Vector bosons 
could also play this role, but would demand an enlargement of the 
MSM symmetry. We have chosen to adopt the simple scalar picture, 
and it is in this direction that we develop the remainder of this 
work.

\section{ECQ beyond standard model}
\label{sec2}

Guided by the procedure employed in section~\ref{sec1}, we will  
investigate appropriate MSM extensions on the light of ECQ. Some 
attempts were already considered where new neutral fermion 
singlets~\cite{marek} or a second Higgs doublet are added to the 
MSM~\cite{zralek}.  A catalogue of baryon number violating scalar 
interactions was also considered in Ref.~\cite{BNV}. However, it 
is interesting to remark that MSM modifications in the direction 
of eliminating the hidden symmetries, $U(1)_{L_e-L_\mu , 
L_e-L_\tau ,L_\mu-L_\tau} $, by performing simple additions in 
its scalar sector, like the inclusion of singlet scalars, singly 
and doubly charged, were not considered yet. Of course, these 
scalars allow some non-standard leptonic interactions and there 
was little experimental motivation for such endeavor, except for 
neutrino physics. In view of the new experimental results related 
to the $(g-2)_\mu$, we are going to limit our study of ECQ to 
extensions which modify only the scalar sector and could equally 
offer an explanation to the discrepancy between the theoretical 
and experimental results on $(g-2)_\mu$~\cite{brown}. 

As briefly pointed in the end of last section, if one wishes to 
break the lepton flavor symmetry combinations, $U(1)_{L_e-L_\mu\, 
,\,\, L_e-L_\tau \,,\,\,L_\mu-L_\tau}$, it is natural to look for 
operators composed by fermions and scalars involving the bilinear 
product, $\bar \Psi^c_i \Psi_j$, which properly accommodates 
family number violation interactions. Here $\Psi$ is a fermion in 
a doublet or singlet representation, the indexes $i$ and $j$ 
denotes the family, and the superscript $c$ means charge 
conjugation. In what follows, we consider two kinds of 
interaction among fermions demanding different species of scalars.

\subsection{lepton-lepton interactions}
\label{sub1}

Within the fermionic representation content of the MSM the 
possible bilinear products involving leptons only, are:
\begin{eqnarray} 
&&\overline{(L_{i_L})^c} L_{j_L} \sim (\mbox{{\bf 1, 
1 $\oplus$ 3},$-(Y_{i_L}+Y_{j_L})$}),\nonumber \\
&&\overline{(l_{i_R})^c} l_{j_R}\sim (\mbox{{\bf 1, 
1},$-(Y_{i_l}+Y_{j_l}$))}. \label{poss}
\end{eqnarray} 

The first term in Eq.~(\ref{poss}) requires either a singlet  or 
a triplet of scalars, both carrying a net total lepton number, 
$L=-2$. Let us first analyze the case of a singlet.  With a 
scalar singlet $h\sim (\mbox{{\bf 1, 1},$Y_h$)}$ we can write  
the following Yukawa interaction,
\begin{equation}
{\cal L}^Y_h = f_{ij} \overline{(L_{i_L})^c} L_{j_L}h + h.c.\,, 
\label{h}
\end{equation}
where $f_{ij}$ is a component of an anti-symmetric $(3\times 3)$ 
family mixing matrix. The above Lagrangian gives us the classical 
constraint relation among the hyper-charges,  
\begin{equation}
Y_{i_L} +Y_{j_L} +Y_h=0\,. \label{hyph}
\end {equation}
The constraints in Eq.~(\ref{hyph}) and Eq.~(\ref{clasconst2}),  
lead automatically to Eq.~(\ref{clasconst4}) which, together with 
Eq.~(\ref{anomcanc}), imply the expected ECQ, assigning to $h$ 
the correct electric charge, $Q_h = 1$. 

The second possibility allowed by the first term in 
Eq.~(\ref{poss}) involves a scalar triplet, $\Delta\sim 
(\mbox{{\bf 1, 3},$Y_\Delta$)}$, composing the following Yukawa 
interaction with the lepton doublets,
\begin{equation}
{\cal L}^Y_h = g_{ij} \overline{(L_{i_L})^c} \Delta L_{j_L} + 
h.c.\,, \label{delta}
\end {equation}
with $g_{ij}$ symmetric. This gives us the subsequent relations 
among the hypercharges, 
\beqq Y_{i_L} +Y_{j_L} +Y_\Delta=0\,, \label{hyp1} \eeqq
which, together with Eq.~(\ref{clasconst2}) and  
Eq.~(\ref{anomcanc}), also result in ECQ. 

Some comments are in order here. This scalar triplet is popular 
in literature and has the following particle content,
\begin{eqnarray} 
\Delta=
\left (
\begin{array}{lcr}
\Delta^0 & \frac{\Delta^+}{\sqrt{2}} \\
\frac{\Delta^+}{\sqrt{2}} & \Delta^{++}
\end{array}
\right ).
\label{triplet} 
\end{eqnarray}
If we allow its neutral component, $\Delta^0$, to develop a 
vacuum expectation value (VEV) the (total) lepton number is 
spontaneously broken through its potential. The main consequences 
are that neutrinos acquire a mass at tree level and a Majoron 
arises. However, such a Majoron is already excluded by 
experiments and it has to be avoided. To accomplish  this we 
assume that $\Delta^0$ does not develop a VEV, which is not a 
fine tuning since this is an equally possible solution to the 
extremum equation that comes from demanding a minimum for the 
potential. In order to clearly see this, let us write down the 
potential involving the standard Higgs doublet, $H$, and the 
triplet, $\Delta$,
\begin{eqnarray}
V(\Phi,\Delta)&=&\mu_H^2 H^{\dagger} H + \lambda_1(H^{\dagger} 
H)^2 
+ \mu_\Delta^2tr(\Delta^{\dagger} \Delta) + 
\lambda_2[tr(\Delta^{\dagger} 
\Delta)]^2 
+\nonumber \\
&&\lambda_3 H^{\dagger} H tr(\Delta^{\dagger} \Delta)+ \lambda_4 
tr(\Delta^{\dagger} \Delta \Delta^{\dagger} \Delta)+ \lambda_5 
(H^{\dagger}\Delta^{\dagger} \Delta H).
\label{potential}
\end{eqnarray}
From the minimum condition over this potential we obtain,
\begin{eqnarray}
&&v_H(\mu_H^2 +\lambda_1v^2_H +\frac{1}{2}(\lambda_3 + 
\lambda_5)v^2_\Delta)=0,\nonumber \\
&&v_\Delta(\mu_\Delta^2 +\lambda_2v^2_\Delta 
+\frac{1}{2}(\lambda_3 + 
\lambda_5)v_H^2 + \lambda_4 v^2_\Delta)=0.
\label{tadpole}
\end{eqnarray}
According to the second relation above, one can promptly observe 
that $v_\Delta =0$ is an equally proper solution for the system 
above. 

Since $\Delta^0$ does not develop a VEV, the total lepton number 
symmetry is kept intact and there will be no mixing among the 
particles of the triplet with those of the doublet. This means 
that the Goldstones in the theory (longitudinal components of 
$W^\pm$ and $Z$) come solely from the Higgs doublet, and the 
masses of the scalars that form the triplet, $M_\Delta$, depend 
only on  $\mu_\Delta$ and $v_H$.  We can safely assume 
$\mu_\Delta \simeq v_H$, which sets $ M_\Delta$ in the 
electro-weak scale, avoiding lower bounds over $\mu_\Delta$. This 
choice will be convenient when discussing the contribution of 
Eq.~(\ref{delta}) to $(g-2)_\mu$ in section~\ref{sec3}.

The last possibility, which stems from the second term in 
Eq.~(\ref{poss}), requires a scalar singlet, $k\sim (\mbox{{\bf 
1, 1},$Y_k$)}$, interacting only with the charged lepton singlets,
\begin{equation}
{\cal L}^Y_k = h_{ij}\overline{(l_{i_R})^c}l_{j_R} k + h.c.\,, 
\label{k++}
\end{equation}
where the coupling $h_{ij}$ is symmetric. The interaction above 
gives us the following relations among the hypercharges, 
\beqq Y_{i_l} +Y_{j_l} +Y_k=0\,,
\label{hyp3}
\eeqq 
which, together with Eq.~(\ref{clasconst2}) and  
Eq.~(\ref{anomcanc}), also implies the ECQ. Once having the ECQ 
we see that the scalar involved in Eq.~(\ref{k++}) carries two 
units of electric charge, $k\equiv k^{++}$. 

\subsection{lepton-quark interactions}
\label{sub2}

Another interesting possibility to include in this picture refer 
to interactions like, $\bar \Psi^c \Psi$, where one of the 
fermions is a lepton and the other a quark. However, the nature 
of the scalar interacting with these fermions is a little subtle, 
requiring that it carries both, barion and lepton charges. 
Scalars like this are known in literature as scalar lepto-quarks. 
Any kind of interaction involving scalar lepto-quarks leads to 
the ECQ and also gives contributions to $(g-2)_\mu$. The fermion 
representation content within the MSM is such that different  
kinds of scalar lepto-quarks are allowed. In general, their 
interactions are classified by $F=0$  and $F=-2$, where $F$ is 
their assigned fermion number. However the kind of bilinear 
fermion product we are interested in here, leaves no room for 
lepto-quark interactions with $F=0$, though they can also drive 
to ECQ~\cite{BNV}. Therefore, the remaining terms in the 
lepto-quark Yukawa Lagrangian are $F=-2$ 
interactions~\cite{cheung}, 
\beq {\cal L}_{F=-2} = g_L \overline{(Q_L)^c} i\tau_2 L_L {\cal 
S}^L_0 + g_R \overline{(u_R)^c} l_R{\cal S}_0^R + \tilde{g_R} 
\overline{(d_R)^c} l_R \tilde{{\cal S}}^R_0 + g_{3L} 
\overline{(Q_L)^c} i\tau_2 {\vec \tau} L_L {\vec{\cal S}}^L_1 
+h.c.\,, \label{leptquar} 
\eeq
where ${\cal S}^L_0 ,\,{\cal S}_0^R ,\,\tilde{{\cal S}}^R_0 $  
are singlets, while $\vec{{\cal S}}^L_1$ is a triplet.  Despite 
the several interaction terms in Eq.~(\ref{leptquar}), in what 
concerns ECQ, just one of them would be sufficient. To be 
convinced of this, notice that any of the terms above connects 
the hyper-charges of leptons with the hyper-charges of quarks by, 
\beqq Y_{quarks} +Y_{leptons}+Y_{{\cal S}} =0\,, \label{hyp4} 
\eeqq
which, together with Eq.~(\ref{clasconst2}) and  
Eq.~(\ref{anomcanc}), lead to the ECQ.

\section{the $(g-2)_\mu$ deviation}
\label{sec3}

The proposals studied in section~\ref{sec2} are very appealing 
from the theoretical point of view, since they deal with simple 
modifications inside the, yet unknown, scalar sector of MSM, 
pointing to a solution for the ECQ problem. At first sight, some 
of them could only add to the list of those models already 
present in literature. However, it is our main goal in this work 
to conciliate those scenarios with the recent $(g-2)_\mu$ 
measurement. Our intent is to show that except for the singlet 
extension, all the previous ECQ solutions can explain the posed 
deviation on $(g-2)_\mu$ with an adequate choice of the 
parameters. This, by itself, would be a strong phenomenological 
motivation to suggest such economic modifications to MSM.

\begin{figure}
\centerline{ \epsfxsize=0.30\textwidth
\begin{picture}(100,240)(-45,50)
\put(-101,165){\makebox(0,0)[br]{a)}} 
\put(-58,155){\makebox(0,0)[br]{$\mu^-$}} 
\put(27,155){\makebox(80,0)[br]{$\mu^-$}} 
\put(40,195){\makebox(80,0)[br]{$S^+\,,\,\,S^{++}$}} 
\put(20,235){\makebox(20,235)[br]{$\gamma$}} 
\ArrowLine(-85,165)(-35,165) \ArrowLine(65,165)(-35,165) 
\ArrowLine(65,165)(115,165) \DashArrowArc(15,165)(50,0,180)5 
\Photon(15,215)(65,240)3 5
\put(-101,85){\makebox(0,0)[br]{b)}} 
\put(-58,75){\makebox(0,0)[br]{$\mu^-$}} 
\put(27,75){\makebox(80,0)[br]{$\mu^-$}} 
\put(10,115){\makebox(80,0)[br]{$S^{++}$}} 
\put(20,60){\makebox(20,60)[br]{$\gamma$}} 
\ArrowLine(-85,85)(-35,85) \ArrowLine(65,85)(-35,85) 
\ArrowLine(65,85)(115,85) \DashArrowArc(15,85)(50,0,180)5 
\Photon(15,85)(65,60)3 5
\end{picture}}
\caption{General charged scalar, $S$, contributions to the muon 
anomalous magnetic moment $(g-2)_\mu$.} \label{fig1}
\end{figure}
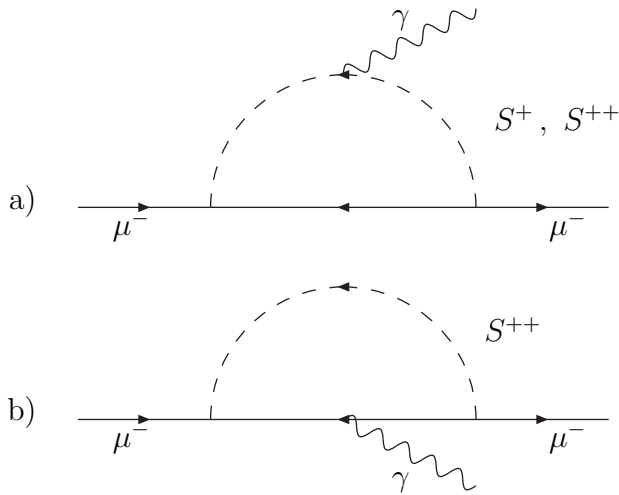

Let us start by situating the $(g-2)_\mu$  problem, originated 
from a new measurement by the BNL experiment~\cite{brown}. It 
indicates a deviation from the theoretical value of 2.6 sigma, 
\beqq a^{exp}_\mu - a^{SM}_\mu = 426\pm 165 \times 10^{-11}. 
\label{deviation} \eeqq
If this result persists~\cite{Yndu} it implies an exciting window 
requiring new physics beyond MSM. Among various scenarios 
proposed to account for the deviation, we restrict ourselves to 
those directly related to extensions in the scalar sector 
discussed in section~\ref{sec2}~\cite{cheung,ps,datta}. As 
observed in Ref.~\cite{ps}, MSM extensions in the scalar sector 
have been almost neglected, mainly because Higgs contributions to 
$(g-2)_\mu$ can be significant only for light masses with usual 
values for Yukawa couplings~\cite{2HDM,miscelania}, namely, 
\beqq f_{\mu \mu} \bar{\mu} \mu H, \label{yakawa1} \eeqq
gives the following contribution to $(g-2)_\mu$~\cite{moore},
\beqq a^\mu_H =\frac{f_{\mu \mu}^2 m_\mu^2}{12 \pi^2 m^2_H}. 
\label{hcont} \eeqq
Here, $f_{\mu\mu}$ is the usual Yukawa coupling for the muon, and 
has the following form
\beqq f_{\mu \mu} = \frac{m_\mu}{v_w} , \label{fmumu1} \eeqq
where, $m_\mu = 0.105$~GeV, is the muon mass and $v_w = 247$~GeV 
is the VEV of the scalar doublet in the MSM. These lead to,
\beqq f_{\mu \mu} \sim 10^{-3}. \label{fmumu2} \eeqq
With this value for $ f_{\mu \mu}$ and considering the Higgs mass 
of the order of hundreds of GeV, $m_H \sim 10^2$~GeV, the standard 
Higgs contribution to $(g-2)_{\mu}$ is negligible,
\beqq a^\mu_H \sim 10^{-13}. \label{smhc} \eeqq
Hence, if we wish to make minimal extensions mimicking this 
sector in order to explain the $(g-2)_\mu$ deviation, within a 
reasonable mass scale for the scalars, we have to impose some 
enhancement over the Yukawa couplings. This is, of course, an 
analysis which could be generalized to other scalar extensions, 
though in the case of lepto-quarks we still can have small 
couplings for fairly large scalar masses~\cite{cheung}. The 
contribution to $(g-2)_\mu$ involving charged scalars is 
diagrammatically depicted in figure~\ref{fig1}.
These pictures can be cast, respectively, in our particular case 
by the following expressions,
\beq &&a)\;\;\;\;a^{S}_\mu = \frac{-C^2 Q_S}{4\pi^2} \int_0^1 
\frac{x^3 -x^2}{x^2 + (z-1)x +\frac{m_f^2}{m_\mu^2}(1-x)} ,
\nonumber \\
&&b)\;\;\;\; a^{S}_\mu = \frac{C^2}{4\pi^2} \int_0^1 \frac{x^2 
-x^3}{x^2 + z(1-x)}, \label{H+H++} \eeq
where $Q_S$ is the scalar charge, multiple of the positive 
electron charge $|e|>0$, $z$ is the ratio between the charged 
scalar and muon masses, $z=\frac{m^2_S}{m^2_\mu}$, and $C$ is a 
factor which involves the matrix coupling as well as a symmetry 
factor for each case (for $i=j$ this symmetry factor is 2, and it 
is 1 otherwise). In the second formula above in 
Eq.~(\ref{H+H++}), it is implicit that we are considering only 
diagonal family interaction for the doubly charged scalar.

Let us first analyze the simplest case given by Eq.~(\ref{h}), 
where we added to MSM only a singlet scalar interacting with 
fermions that, after assigning to it the correct electric charge, 
becomes 
\begin{equation}
{\cal L}^Y_{h} = f_{ij} \overline{(L_{i_L})^c} L_{j_L}h^+ + h.c. 
\,. \label{sing+}
\end{equation}
We recall that the coupling $f_{ij}$ is anti-symmetric on the 
family indices ($i$ and $j$), and are roughly constrained if one 
considers $e-\mu-\tau$ universality~\cite{babu}, yielding
\beqq \frac{f_{e\mu}^2}{m_h^2}\lesssim 8\times 
10^{-8}~\mbox{GeV}^{-2}\,, \label{fcons}
\eeqq
which can be translated into $f_{e\mu}\lesssim 10^{-2}$ for 
$m_h\sim 100$~GeV (similar constraints can be imposed on 
$f_{e\tau}$ and $f_{\mu\tau}$).

With this value for the Yukawa coupling we have $C=0.5\, 
f_{ij}=5\times 10^{-3}$, with $i=\mu$ and $j=e\,,\tau$ in the 
first expression of Eq.~(\ref{H+H++}). Since the interaction 
involves a singly charged scalar, the fermion inside the loop has 
to be a neutrino, which is assumed massless in this work, so 
$m_f=m_\nu=0$ in this case. However, in this regime we are not 
able to get any significant contribution to $(g-2)_\mu$ for 
reasonable scalar masses. We observe that this singly charged 
scalar can participate in more complex models, like the extended 
Zee-model, where a second higgs doublet, a second singly charged 
scalar and a right-handed neutrino are added along with a new 
$U(1)$ symmetry. These models aim to explain neutrino mass 
through radiative corrections and can lead to a substantial 
effect on $(g-2)_\mu$~\cite{zeegmu} when the scalar mass is 
between $100<M_S<300$~GeV. Although this alternative to the 
singlet singly charged scalar alone fits well the scenario we 
have in mind, once the inclusion of such additional particles do 
not affect the achieved ECQ, it is less appealing concerning its 
complexity.  

The second case to be studied is the triplet one, with particle 
content given by Eq.~(\ref{triplet}). As we have seen in 
section~\ref{sec2}, we can avoid phenomelogical complications by 
taking the alternative solution for the VEV of this triplet to be 
zero. Still, we can assume its mass scale to be of the same order 
of Electro-Weak symmetry breaking scale, $M_\Delta\sim 200$~GeV. 
In this range the interaction term in Eq.~(\ref{delta}) gives the 
appropriate contribution to $(g-2)_\mu$. In order to see this 
observe that the interaction with singly and doubly charged 
scalars in Eq.~(\ref{delta}) can be written explicitly as, 
\begin{equation}
{\cal L}^{\prime Y}_{\Delta} = 
g_{ij}\{\,[\overline{(\nu_{i_L})^c} l_{j_L} 
+\overline{(l_{i_L})^c} \nu_{j_L}]\Delta^+ 
+\overline{(l_{i_L})^c} l_{j_L} \Delta^{++}\,\} + h.c.\,. 
\label{deltacont}
\end{equation}
Here, $g_{ij}$ is symmetric on the family indices, $i$ and $j$.  
Clearly what matters in solving the ECQ is the very property of 
such interactions to violate lepton flavor conservation, but in 
what concerns $(g-2)_\mu$, these violating terms are suppressed. 
This happens essentially because we are assuming scalars with 
mass on the Electro-Weak symmetry breaking scale. For instance, 
consider only three of the flavor changing process: $\mu 
\rightarrow 3e$, $\tau\rightarrow 3\mu\,,\,\,3e$. The decay rate 
of a lepton, $l^{\prime}$, in three lighter leptons, $l$, allowed 
by the interaction in Eq.~(\ref{deltacont}) has, in general, the 
following expression~\cite{babu}, 
\beq \Gamma(l^{\prime} \rightarrow 3l)\simeq 
\frac{g^2_{l^{\prime} l} 
g^2_{ll}}{192\pi^3}\frac{m^5_{l^{\prime}}}{m^4_\Delta}. 
\label{rate} \eeq
The present experimental bounds on these flavor changing 
processes are: $BR(\mu \rightarrow 3e) \lesssim 10^{-12}$, 
$BR(\tau \rightarrow 3e, 3\mu) \lesssim 10^{-6}$\cite{pdg}. These 
can be translated to the following constraints: $\frac{g_{e \mu} 
g_{ee}}{m^2_\Delta}\lesssim 10^{-11} $~GeV$^{-2}$ and $\frac{g_{e 
\tau} g_{ee}}{m^2_\Delta}, \frac{g_{\mu \tau} g_{\mu 
\mu}}{m^2_\Delta} \lesssim 10^{-7}$~GeV$^{-2}$. If we have a 
scalar with mass, $m_\Delta \simeq 10^2$~GeV, such constraints 
require: $g_{e\mu}g_{ee}\lesssim 10^{-7}$ and 
$g_{e\tau}g_{ee},\,\,g_{\mu\tau}g_{\mu\mu} \lesssim 10^{-3}$. 
Concerning the diagonal components, there is a lower bound to the 
product of $g_{ee}$ and $g_{\mu\mu}$ imposed by 
muonion-antimuoniun conversion: $\frac{g_{ee}g_{\mu 
\mu}}{m^2_\Delta}>10^{-8}$~GeV$^{-2}$~\cite{mohabook}. For 
$m_\Delta \simeq 10^2$~GeV we have $g_{ee}g_{\mu \mu}>10^{-4}$. 
Along with this the only upper bounds come from $(g-2)_e$ and the 
Bhabha scattering process. From these,  the last is the most 
stringent~\cite{bhabha}: $\frac{g_{ee}^2}{m^2_\Delta} 
<10^{-6}$~GeV$^{-2}$, which requires, in our case, $g_{ee}< 
10^{-1}$. There is no experimental constraint on $g_{\mu \mu}$, 
except for the recent $(g-2)_\mu$ deviation, which we are going 
to expose below. 

Since we are interested in enhanced diagonal Yukawa couplings, it 
is clear that the off-diagonal ones are suppressed by, at least, 
three orders of magnitude. In this context, we can safely assume 
$g_{\mu\mu}\simeq 1$ and, compute $(g-2)_\mu$ to set the suitable 
mass range for the scalar, $\Delta$, which would render agreement 
between theory and experiment. In the triplet case, both 
expressions in Eq.~(\ref{H+H++}) have to be employed. The first 
of these expressions accounts for a $\Delta^+$ as well as a  
$\Delta^{++}$ exchange, while the second only involves 
$\Delta^{++}$. The singly charged scalar contribution is similar 
to the singlet case, except for the Yukawa coupling, which is 
diagonal in this case and of the order of unity, leading to 
$C=0.5$ for this computation. For the doubly charged scalar 
exchange, $C=1$ at both expressions in Eq.~(\ref{H+H++}). In 
order to explain the observed deviation in $(g-2)_\mu$ we have to 
keep the triplet mass between $200$~GeV$\lesssim m_\Delta 
\lesssim 300$~GeV, which makes it an attractive proposal, not 
only for solving simultaneously the ECQ puzzle, but for it forces 
a mass range which could be easily prompted in the next 
generation of accelerators.

We consider now the doubly charged scalar singlet, 
Eq.~(\ref{k++}). This extension was already studied in 
Ref.~\cite{ps}, wherein it was verified that such a scalar is 
relevant for the $(g-2)_\mu$ problem, only if its mass is around 
$200$~GeV. This is not surprising, since the triplet case studied 
above only differ from this case by a singly charged scalar 
contribution, which is not as important as the doubly charged one 
because the former is almost one order of magnitude lesser than 
the last. Hence, this scenario is an equally  good candidate to 
be the solution for both, ECQ and $(g-2)_\mu$ problems. 

Finally, we can discuss the role of scalar letpo-quark 
interactions, as given by Eq.~(\ref{leptquar}), in the context of 
$(g-2)_\mu$, assessed in Refs.~\cite{cheung,datta}. In 
Ref.~\cite{datta}, the mixing among generations was allowed, 
which could lead to problems concerning flavor changing neutral 
currents. This was avoided in Ref.~\cite{cheung} by assuming 
there was no such generation mixing and the author obtained that 
the only important effect on $(g-2)_\mu$ occurs when the 
lepto-quark has both, left and right handed couplings to 
fermions, although lepto-quarks coupling with only one type of 
handedness can harmlessly coexist. We will adopt this approach 
here. In our case, these lepto-quarks are identified in the 
Yukawa Lagrangian, Eq.~(\ref{leptquar}), as ${\cal S}^{L,R}_0$.  
The bounds put by $(g-2)_\mu$ over their masses are,  
$0.7$~TeV$<M_{\cal{S}}<2.0$~TeV, and were obtained considering 
couplings of electromagnetic size~\cite{cheung}.

In this context, it is still possible to assume first-second 
generation universality or not. The first possibility puts 
additional constraints on the allowed lepto-quark interaction 
through $\eta_{CC}$\footnote{$\eta_{CC}=\eta^{ed}_{LL} - 
\eta^{eu}_{LL}=(0.051 \pm 0.037)$ TeV where $\eta^{lq}_{\alpha 
\beta}$ is a contact parameter, with $\alpha$ and $\beta$ 
denoting the chirality of the lepton ($l$) and the quark ($q$), 
used to parameterize the four fermion effective interaction that 
appear in a regime where the mass of the lepto-quarks is larger 
than the energy scale involved in the experiment.}~\cite{cheung}. 
To avoid this, the singlet scalar lepto-quark has to be 
accompanied by the triplet, which does not significantly 
contributes to $(g-2)_\mu$ while modifying $\eta_{CC}$ in the 
right direction such as to compensate the effect of the singlet. 
The second possibility alone is enough to circumvent this 
complication since it leads to the desired effect on $(g-2)_\mu$ 
without requiring the presence of other lepto-quarks. As the 
coexistence of additional lepto-quarks does not jeopardize our 
picture, both possibilities are welcome, though the second is 
more economic. 

\section{conclusions}
\label{sec4}

In this work we have suggested small extensions of the MSM, by 
augmenting the scalar sector on minimal portions, aiming to 
explain both, the long-standing problem of electric charge 
quantization and the measured deviation of the muon anomalous 
magnetic moment, recently reported by the BNL experiment. The 
scalars suggested here are not usual ones though, since they must 
couple to bilinear fermionic products, $\bar \Psi^c \Psi$, and 
consequently carry two units of lepton number. The purpose is to 
oblige the new Yukawa interactions to explicitly induce family 
mixing, eliminating the hidden symmetries which impede the 
realization of ECQ inside the MSM. Notice that such extensions do 
not interfere with other symmetries of the MSM, keeping them 
intact. 

Among the proposed scalars there are singlets, doubly and singly 
charged, a single triplet, and also lepto-quarks, disposed in 
simple configurations. All of them are fair candidates to 
simultaneously achieve ECQ and an explanation for the 
$(g-2)_\mu$, except for the singly charged singlet, which  
accounts for ECQ but it is insufficient to properly solve the 
$(g-2)_\mu$ discrepancy. This singlet could be inserted in a more 
complex configuration in order to accomplish this double task, 
although we would rather stick to plain extensions of MSM. It is 
interesting to remark that the scalar masses rendered by the this 
study are close to the experimental reach of the next generation 
of accelerators. 

In summary, we successfully managed to relate some solutions for 
the ECQ with the theory-experiment deviation on the muon 
anomalous magnetic moment. The whole picture impel us to suggest 
that agreement between experiment and theory are, together, 
pointing the direction to follow, which in this situation is a 
modest expansion of the MSM scalar sector. 
 

{\it Acknowledgements.}  Work supported by Funda\c c\~ao de Amparo 
\`a
Pesquisa do Estado de S\~ao Paulo
(FAPESP).

\end{document}